\documentclass[]{raa}
\usepackage{graphicx}
\usepackage{times}
\usepackage{natbib}
\usepackage{mathrsfs}
\usepackage{amsmath}
\usepackage{amssymb}
\usepackage{booktabs}
\usepackage{hyperref}

\begin{document}

   \title{The optical/UV excess of X-ray-dim isolated neutron star. II. Nonuniformity of plasma on strangeon star surface
}
   \volnopage{Vol. 000 No.0, 000--000}      
   \setcounter{page}{1}          

   \author{Weiyang Wang
      \inst{1,2,3}
   \and Yi Feng
      \inst{4,5,6}
   \and Xiaoyu Lai
      \inst{7}
      \and Yunyang Li
       \inst{3}
      \and Jiguang Lu
       \inst{4,8}
   \and Xuelei Chen
       \inst{1,2,9}
 \and Renxin Xu
       \inst{3,10}
   }

   \institute{Key Laboratory of Computational Astrophysics, National Astronomical Observatories, Chinese Academy of Sciences, Beijing 100012, China {\it wywang@bao.ac.cn}\\
        \and
            School of Astronomy and Space Sciences, University of Chinese Academy of Sciences, Beijing 100049, China\\
        \and
          School of Physics and State Key Laboratory of Nuclear Physics and Technology, Peking University, Beijing 100871, China\\
          \and
National Astronomical Observatories, Chinese Academy of Sciences, Beijing 100012, China
           \and
         CAS Key Laboratory of FAST, National Astronomical Observatories, Chinese Academy of Sciences, Beijing
          \and
          School of Astronomy and Space Sciences, University of Chinese Academy of Sciences, Beijing 100049, China\\
        \and
        Hubei University of Education, Wuhan 430205, China
 \and
 Key Laboratory of Radio Astronomy, Chinese Academy of Science, Beijing 100012, China
           \and
Center for High Energy Physics, Peking University, Beijing 100871, China
\and
Kavli Institute for Astronomy and Astrophysics, Peking University, Beijing 100871, China
   }

\date{Received~~2017 October 16; accepted~~2017~~Feb 22}

\abstract{X-ray-dim isolated neutron stars (XDINSs), also known as the Magnificent Seven, exhibits a Planck-like soft X-ray spectrum. In the optical/ultraviolet(UV) band, there is an excess of radiation compared to the extrapolation from the X-ray spectrum.
However, the majority exhibits ``spectral deviations'': the fact that there are more flux at longer wavelengths makes spectra deviating from Rayleigh-Jeans laws.
A model of bremsstrahlung emission from a nonuniform plasma atmosphere is proposed in the regime of a strangeon star to explain the optical/UV excess and its spectral deviation as well as X-ray pulsation.
The atmosphere is on the surface of emission-negligible strangeon matter, formed by the accretion of ISM-fed debris disk matter moveing along the magnetic field lines to near polar caps, and these particles may spread out of the pole regions that makes the atmosphere non-uniform. The modeled electron temperatures are $\sim100-200$\,eV with radiation radii $R_{\rm opt}^{\infty}\sim5-14$\,km.
The spectra of five sources (RX J0720.4--3125, RX J0806.4--4123, RX J1308.6+2127, RX J1605.3+3249, RX J1856.5--3754) from optical/UV to X-ray bands could be well fitted by the radiative model, and exhibit gaussian absorption lines at $\sim 100-500$\,eV as would
be expected. Furthermore, the surroundings (i.e., fallback disks or dusty belts) of XDINSs could be tested by future infrared/submillimeter observations.}


\authorrunning{W. Y. Wang et al.}

\titlerunning{The optical/UV excess of XDINS}
\maketitle

\section{Introduction}
The Magnificent Seven refers to seven X-ray-dim isolated neutron stars (XDINSs) with thermal radiation of soft X-rays, which offer an unprecedented opportunity to unveil their surface temperature and magnetic field as well as the state of dense matter at supra nuclear densities \citep{tur09}.
They are located in the upper right corner of $P-\dot P$ diagram together with a few high-$B$ pulsars and magnetars, indicating that they may have strong magnetic fields \citep{mor03,ton16}. 
These objects,  RX J1856.5--3754(J1856 hereafter as other sources) for instance, are characterized by Planck-like spectra (no power law component) in the X-ray band with a relatively steady flux over long timescales  \citep{bur01,ho07,van07,kap09}.
Compared with the extrapolated blackbody X-ray spectrum, the XDINS exhibits an excess in optical/UV flux, with a factor of $\sim5-50$\citep{kap11}.
However, it is {\it not an alleged} observation that the optical flux appears to deviate from a R-J distribution. The spectrum can be described as a power law emission ($F_{\nu}\varpropto\nu^{\beta}$). For R-J spectrum $\beta=2$, but the spectra for XDINS are flatter \citep{kap11} with the only exception of RX J0420.0--5022.
Also noteworthy is the fact that the spectral deviations of most XDINSs are greater than the photoelectric absorptions by neutral hydrogen as modeled from X-ray spectra.
Certainly, more powerful observations would make sure if the deviations are real, but we are focusing to understand the flat spectrum here.

To explain the optical excess, in the regime of conventional neutron star (NS), authors proposed two representative models.
(1). \cite{tru04} and \cite{tur04} adopted a model of ``bare'' (i.e., no gaseous atmosphere sits on the top of the crust) NS which spectrum can be described as a two-component blackbody. In the case of J1856 (at a distance $d \sim 120$\,pc from us), for example, the temperature of the hot spot is $kT_{\rm X}^\infty \sim 63.5\,\mathrm{eV}$ with radiation radius $R_{\rm X}^\infty \sim 4.4\,(d/120\,\mathrm{pc})\,\mathrm{km}$, while the cold part has $kT_{\rm opt}<33\,\mathrm{eV}$ and $R_{\rm opt}^\infty> 17\,(d/120\,\mathrm{pc})\,\mathrm{km}$ \citep{bur03,tru04}.
They are fine tuning the geometry in order to explain the small X-ray pulsed fraction (PF) observed, PF\,$<1.3\%$.
Recently, the gravitational wave observation of GW 170817 provides valuable constraints of tidal deformability, $\Lambda(1.4M_\odot)<10^3$ \citep{abb17}.
According to this result, \cite{ann17} concluded that the maximal radius is $\sim14$ km if the sound speed of NS is smaller than $c/\sqrt{3}$, with $c$ the speed of light.
Therefore, a NS with large radius ($>17$ km) is not acceptable.
(2). \cite{ho07} proposed a model of thin, magnetic, partially ionized hydrogen atmosphere on top of a condensed iron surface to fit the spectrum of J1856.
However, a condensed surface at the bottom of atmosphere, brings more low-energy photons so that it has to introduce a large value of hydrogen absorption that would consequently present a strong photoelectric absorption which deviates from the Rayleigh-Jeans (R-J) spectrum of J1856 at optical bands.
Additionally,  both of these two NS models face a common problem that the cohesive energy for iron atoms and molecular chains in strong magnetic fields is quite uncertain since this small number is the difference of two large numbers \citep{Mue84}.

However, the inner structure of pulsar remains unclear, which depends on the challenging problem of fundamental strong interaction at low energy scale. There are thus many speculations, among which a strangeon star model is proposed to understand different manifestation of compact objects \citep{lai17}.
A bremsstrahlung radiative model of a strangeon star atmosphere was proposed to explain the optical/UV excess \citep{wang17}.
The atmosphere is modeled with two temperature components \citep{xu14}, formed and maintained by accretion, and can be simply considered as the upper layer of a normal NS.
The radiation is optically thick at optical bands while optically thin at X-ray bands.
Thus, the observed spectra of the Seven could be well fitted by the radiative model, from optical to X-ray bands, except the spectral deviations from R-J regime at optical/UV bands.
In this paper, we propose that the spectral deviations come from non-uniformities of the atmosphere which is on top of an emission-negligible strangeon star surface.
The fallback disk matter accreted moves along the magnetic field lines to near polar caps, and parts of them may diffuse to other regions,
generating a nonuniform distribution of the atmosphere.
Thus, the spectrum may deviate from a R-J radiation at optical bands because the bremsstrahlung emissivity is particle distribution-dependent (as demonstrated in Fig. \ref{fig1}).

In section 2.1, we interpret that the strangeon matter could be emission-negligible.
In Section 2.2, we propose a toy model to describe the non-uniformity of the atmosphere and show that this distribution could make the spectrum deviating from the R-J regime.
In Section 3, we present the fitting details and results for the X-ray data (Sec. 3.1) and optical data (Sec. 3.2).
Finally, we discuss various issues in this model in Section 4, and make a brief summary in Section 5.

\section{Radiation from a strangeon star}
\subsection{The strangeon star surface}
\begin{figure}
\includegraphics[width=0.47\textwidth]{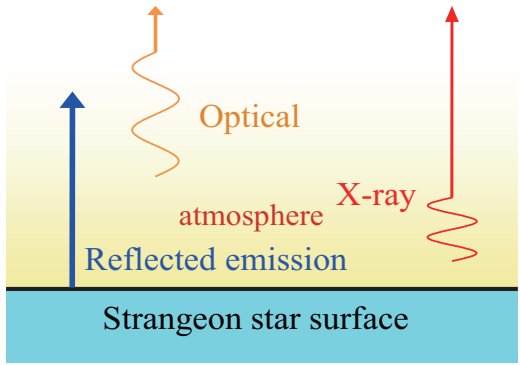}
\caption{\small{A schematic diagram of the observed radiation from the atmosphere which is on top of a strangeon star. The total observed radiation consists of the emission from the atmosphere and the reflected component by the emission-negligible surface. The radiation is optically thin at X-ray bands so that photons can emit from much deeper location, while optically thick for long wavelength photons.}}
\label{fig0}
\end{figure}

The radiation from strangeon matter is totally neglected in the calculation as following, and this could be valid because of the high plasma frequency, $\omega_{\rm p}$.
The plasma frequency of degenerated charged Fermion gas, $\omega_{\rm p}\propto n^{1/3}$, with $n$ the number density~\citep{uso01}.
For strangeon matter with baryon density, $n_{\rm b}$, to be a few nuclear density, the electron number density $n_{\rm e}=(10^{-5}\sim 10^{-6})n_{\rm b}$, with baryon number density $n_{\rm b}=(1.5-2.0)n_{\rm nuc}$ , where $n_{\rm nuc}$ is the normal nuclear matter density \citep{alc86}, and then the frequency $\omega_{\rm p}\sim 100$\,keV is so high that the optical as well as soft X-ray emissivity is negligible.
Therefore, the strangeon star's atmosphere could be similar to the upper layer of a normal NS, but there do exist differences (Fig \ref{fig0}).
The emission-negligible surface does not bring a lot of high energy photons that is coincidence with the fact of the high energy tail absence in the X-ray spectrum of XDINS (e.g., \citealt{vanet07}).

However, the strangeon matter surface exhibit an extremely high reflectance that can be regarded as total reflection.
The reflected component by the strangeon matter surface should certainly be included in the total emission.
Then, the observed flux is multiplied by a factor of $[1+\mathrm{exp}(-\tau_{\infty})]$, where the factor of $\mathrm{exp}(-\tau_{\infty})$ results from the reflection (see Appendix) and illustrated in Figure \ref{fig0}.
The reflection effect intensifies X-ray emissions while can be ignored in optical bands.

\subsection{The plasma atmosphere with non-uniformity: A toy model}\label{sec2.2}
\begin{figure}
\includegraphics[width=0.48\textwidth]{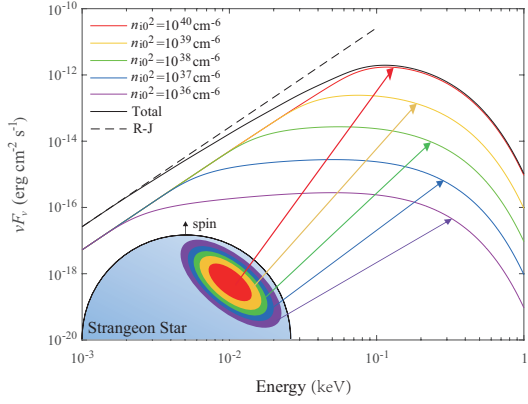}
\caption{\small{The observed spectrum of a rotating NS with a nonuniform atmosphere. The total radiation consists of five components with same radiative areas while different number density: $n_{\rm i0}^2=10^{40}\,{\rm cm}^{-6}$ (red solid line), $n_{\rm i0}^2=10^{39}\,{\rm cm}^{-6}$ (yellow solid line), $n_{\rm i0}^2=10^{38}\,{\rm cm}^{-6}$ (green solid line), $n_{\rm i0}^2=10^{37}\,{\rm cm}^{-6}$ (blue solid line) and $n_{\rm i0}^2=10^{36}\,{\rm cm}^{-6}$ (purple solid line), for given $T_{\rm e}=0.1\,$keV, $T_{\rm i}=0.1$\,MeV and $(R_{\rm opt}^{\infty}/d)^2=(10\,{\rm km}/0.25\,{\rm kpc})^2$ . The total radiation (the black solid line) of these five components exhibits a deviation compared with a R-J spectrum (dotted line).}}
\label{fig1}
\end{figure}

A nonuniform plasma atmosphere could produce a bremsstrahlung radiation spectrum deviating from the R-J law at low energy (optical/UV) bands,
because the optical depth of the  radiation is energy-dependent \citep{wang17}.
The total flux is the sum of radiation from local regions (see Equation (\ref{a3}) in Appendix).
In these cases, a nonuniform atmosphere may enable the spectral index to deviate from a R-J spectrum.
A sample of deviated spectra from a nonuniform atmosphere are shown in Figure \ref{fig1}.
It is worth noting that the radiation in high energy is mainly from the polar caps.
Therefore, the data of both high energy and low energy can be fitted separately (see Section 3).
Actually, the observed flux is the time-averaged radiation from a rotating NS and the detailed calculations are shown in the Appendix.

\begin{figure}
\includegraphics[width=0.47\textwidth]{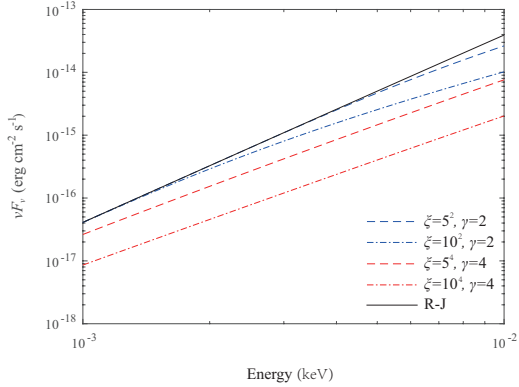}
\caption{\small{Comparison of photo indexes of the spectrum with $\xi=5$, including $\gamma=2$ (bule dashed line) and $\gamma=4$ (red dashed line), and the corresponding case $\xi=10$ with $\gamma=2$ (blue dashed-dotted line) and $\gamma=4$ (red dashed-dotted line). The solid line is a R-J spectrum from a uniform atmosphere.}}
\label{fig2}
\end{figure}

Note that the temperature gradient could not be the reason for the deviation because of the high thermal conductivity of the star surface unless there is an extremely strong magnetic field ($\sim 10^{15}$\,G, e.g., \citealt{pon09}).
Thus, we regard the non-uniformity of the number density as the main reason for the fact of spectral index deviation.
However, a Goldreich-Julian-like distribution \citep{gol69} in a dipolar magnetic field cannot explain the deviation because the regions which have low density radiative matters are very small.
There are some possible reasons for the non-Goldreich-Julian-like distribution: i) some atmospheric matter gyrate around the magnetic lines and diffuse to other regions; \textrm{ii}) there are some multipolar magnetic fields that support particles moving into high latitude regions from polar caps.

As we have discussed in Sec 2.1, the atmospheric matter is accreted by gravity and moves along the magnetic field lines to near polar caps.
However, there may be some multipolar magnetic fields that help small parts of the accreted matters spread out of near polar caps exhibiting a nonuniform particle distribution.
Here, based on the above analysis, it is assumed that the the atmosphere: i) is approximately uniform at polar caps; \textrm{ii}) can be described by a power-law-like distribution at high latitudes; \textrm{iii}) deceases rapidly at the edges of polar caps.
To simply account the complicated atmospheric distribution in the magnetosphere, we propose a toy model to describe the nonuniformity.
An assumed distribution of number density can be constructed,
\begin{equation}\label{8}
n_{\rm e}=n_{\rm i}=n_{\rm i0}(\theta)\mathrm{exp}(\frac{-m_{\rm i}gz}{kT_{\rm i}})=n_0\frac{\mathrm{exp}(-\frac{m_{\rm i}gz}{kT_{\rm i}})}{1+\xi\cdot\theta^{\gamma}},
\end{equation}
where $z$ measures the height above the star's surface, $n_0$ is the number density of ions on the star's surface at $\theta=0$, $m_{\rm i}$ is the mass of a ion(mainly proton), $T_{\rm i}$ is the temperature of ions, $g$ is the gravitational acceleration above the surface of a strangeon star, and $\xi$ as well as $\gamma$ are two constants to describe the nonuniformity, respectively.
The combination term $\xi\cdot\theta^{\gamma}$ is supposed to be one near the polar cap.
In this case, the plasma atmosphere would be approximately uniform if all the ISM-fed debris disk accreted matter can diffuse from the polar caps to other parts of the stellar surface (i.e., $\xi=0$, where a Rayleigh-Jeans spectrum is shown at optical bands).
In order to show the impacts of $\xi$ and $\gamma$ on the spectral index, we calculate the observed time-averaged spectrum with $kT_{\rm e}=0.1\,$keV, $n_{0}^2kT_{\rm i}=10^{42}$\,keV\,${\rm cm^{-6}}$, $(R_{\rm opt}^{\infty}/d)^2=(10\,{\rm km}/0.25\,{\rm kpc})^2$, $\zeta=0$ and $\alpha=0$ (definition of $\zeta$ and $\alpha$, see Appendix).
The comparison of deviation with different $\xi$ and $\gamma$ is shown in Figure \ref{fig2}.

\section{Data Reduction and Fitting}

\subsection{X-ray data}

\begin{table}
\begin{center}
\caption{Summary of Photometry and some parameters for XDINSs}
\begin{tabular}{cccccc}
\hline \hline
Source & Optical (mag) & $\beta$ & PF ($\%$)\\
\hline
J0420 & $B=26.6$ & $2.20\pm0.22$ & 13\\
J0720 & $B=26.6$ & $1.43\pm0.12$ & 8--15\\
J0806 & $B>24$ & $1.63\pm0.20$ & 6 \\
J1308 & $m_{\mathrm{50ccd}}=28.6$ & $1.62\pm0.14$ & 18\\
J1605 & $B=27.2$ & $1.23\pm0.07$ & -- \\
J1856 & $B=25.2$ & $1.93\pm0.08$ & $<$1.3 \\
J2143 & $B>26$ & $0.53\pm0.08$ & 4 \\
\hline \hline
\end{tabular}
\label{tab1}
\end{center}
{\bf Notes.} The optical data quoted form \cite{kap11} of each XDINS can be fitted by a power law ($F_{\nu}\varpropto\nu^{\beta}$, where $\beta$ is the photo index). The optical stellar magnitude and PF shown in this table are quoted form \citealt{hab07}.
\end{table}

The details of X-ray data reduction from {\em XMM-Newton} for the Seven can be seen in \cite{wang17}.
The XDINS spectral analysis is performed with XSPEC 12.9 \citep{arn96}, selecting photon energies in the range of 0.1--1.0\,keV.
To omit some degeneracies, two angles $\zeta$ and $\alpha$ are treated as zero because the spectra are not going to be felt in the range of $\zeta<30^{\rm o}$ and $\alpha<15^{\rm o}$.
In addition, there is little decrease for flux with $\zeta$ and $\alpha$ increasing while not for the spectral index.

The optical and X-ray data can be fitted separately because the emission is optically thin at high-energy bands while optically thick at optical bands.
We first fit X-ray data with the uniform atmosphere model accounting for polar caps mainly emit high energy photons.
The results of spectral fitting by the uniform model with Gaussian absorption are presented in \cite{wang17}.
Therefore, we fix $N_{\rm H}$, $T_{\rm e}$ and $E_{\rm Line}$ as the modelled uniform atmosphere while command $y/R$(where $y=n_{\rm i0}^2kT_{\rm i}$, $R$ is assumed to be $R_{\rm opt}^{\infty}$) and width $\sigma$ as free parameters.
And then we treat all these fixed parameters as free, and extract the best-fit $N_{\rm H}$ to fix photoelectric absorption.

\begin{figure*}
\centering
\includegraphics[width=1\textwidth]{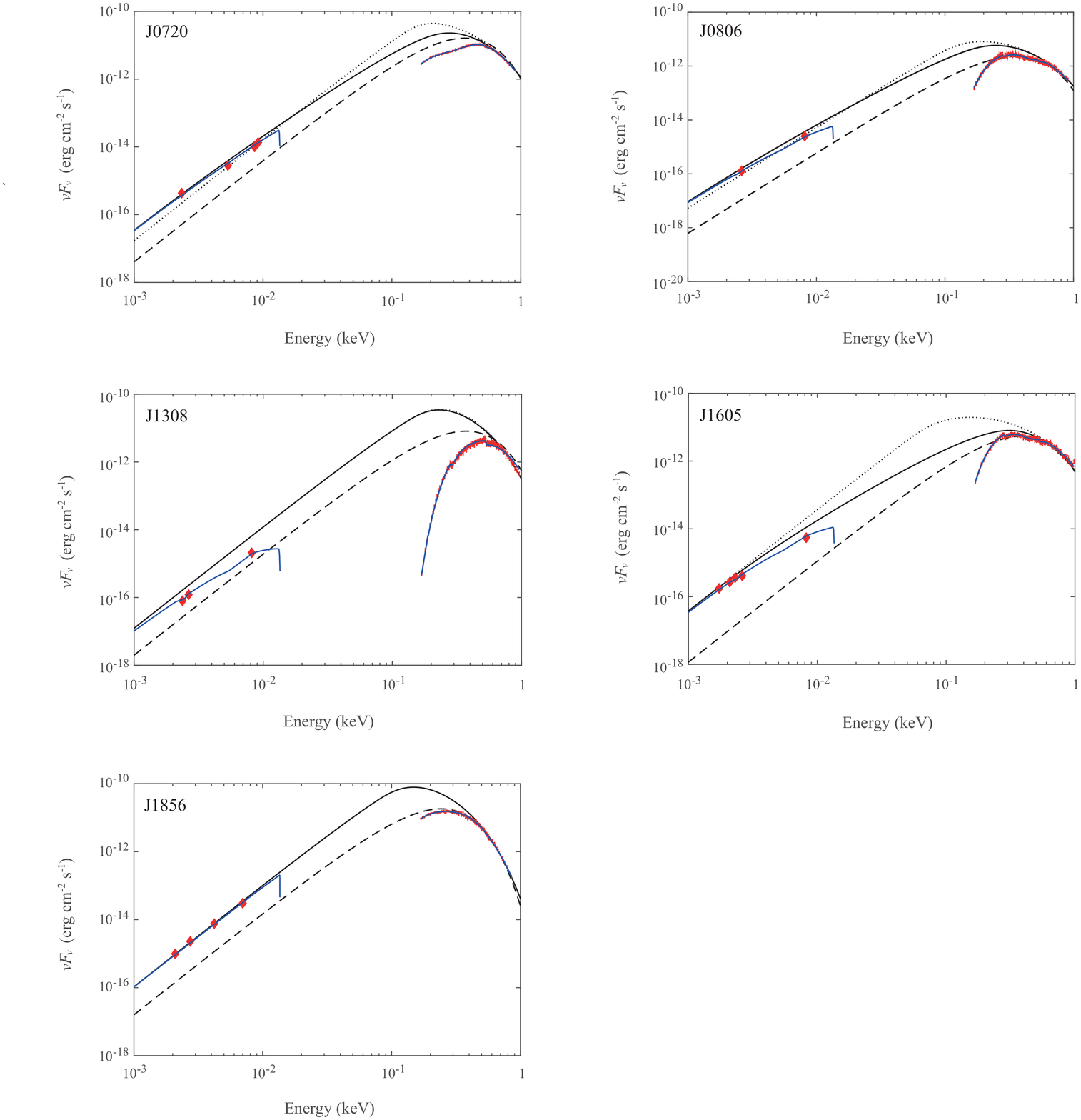}
\caption{\small{The spectra of J0720(top left), J0806(top right), J1308(center left), J1605(center right) and J1856 (bottom) from optical to X-ray bands are shown in this figure. The absorbed best-fit nonuniform radiative model of each source is the blue solid line, while the unabsorbed nonuniform radiative model is the black solid line. To exhibit the optical/UV excesses, pure blackbody (black dashed line) extrapolated from each X-ray spectrum are plotted. The archival X-ray data with their error bars from {\em XMM-Newton} (red dot) and optical data (red diamond) from {\em HST} are also plotted. By comparison, the uniform radiative model (black dotted line) is also plotted.}}
\label{fig3}
\end{figure*}

\subsection{Optical data}

Optical counterparts for each XDINS have been searched in deep optical observations by {\em HST}, and detailed photometry measurements and data errors are presented in \cite{kap11}.
Each photometry spectrum of XDINS can be fitted by a power law and fixed by a function of extinction in which the value is determined by the column density of hydrogen $N_{\mathrm{H}}$ from X-ray spectrum data fitting (details of XSPEC photoelectric absorption are shown in \citealt{mor83}).

To describe a photoelectric absorption when the light transmitting through an extinction layer, the atmospheric extinction in magnitudes $A_{\lambda}$ is defined (see, \citealt{car83}, Equation (3)), and optical extinction $A_V$ depends on $N_{\rm H}$ which is given by X-ray spectral fitting: $A_V=N_{\rm H}/1.79\times 10^{21}\,{\rm cm}^{-2}$ \citep{Pre95}.
In general, the extinction value $A_{\lambda}/A_V$ depends on wavelength and the value of the overall extinction $A_V$.
However, $A_{\lambda}/A_V$ can be considered as wavelength-dependent since the changes are less than $1\%$ in $A_{\lambda}/A_V$ for the range of $N_{\rm H}=(1-4)\times10^{20}\,{\rm cm}^{-2}$ \citep{kap11}.
The absorbed flux can be described as
\begin{equation*}
F_{\nu}=F_{\nu0}\cdot10^{-2.23\frac{A_{\lambda}}{A_V}\cdot\frac{N_{\rm H}}{10^{22}\,{\rm cm^{-2}}}},
\end{equation*}
where $F_{\nu0}$ is the unabsorbed flux.
\cite{van01a} presents a approximate wavelength-dependent values of $A_{\lambda}/A_V$ for an unabsorbed $10^6$\,K($\sim100$\,eV) blackbody.

Then, with fixed photoelectric absorption, the optical data are fitted by the nonuniform radiative models.
In this process, the optical data are fitted by the absorbed nonuniform radiative model with free $R_{\rm opt}^{\infty}$ as well as fixed $\xi$ and $\gamma$.
We list the values of $\xi$ and $\gamma$ that are at $\geq1\sigma$ confidence and the modelled $R_{\rm opt}^{\infty}$ in Table 2.
With these parameters, the radiative spectra and spectral data of J0720, J0806, J1308, J1605 and J1856 are reproduced and plotted in Figure \ref{fig3}.
The X-ray data are also fitted by the blackbody model while only J1856 shows a better fit ($\chi^2=1.11$) than the bremsstrahlung emission.

Therefore, the spectra of J1308 and J1856 can be well fitted by R-J-like emissions which are photoelectric absorbed without nonuniformity.
For J1605, X-ray data fitting with a uniform radiative model presents a large value of $N_{\rm H}\sim3\times10^{20}\,{\rm cm^{-2}}$.
In this case, the absorbed nonuniform emission can well fit the deviate spectrum.
However, the nonuniform radiation predicts that $N_{\rm H}$ would be smaller than $\sim1.5\times10^{20}\,{\rm cm^{-2}}$ because the nonuniformity of the atmosphere decreases the emissions from polar caps in low-energy X-ray bands ($\sim0.1-0.4$\,keV).
The same case for J0806 that $N_{\rm H}$ is estimated to be $\sim2.2\times10^{20}\,{\rm cm^{-2}}$, which is supposed to be smaller than Galactic value $2.7\times10^{20}\,{\rm cm^{-2}}$ obtained from \cite{kal05}.
While J1308 shows a larger value of $N_{\rm H}$ that may indicates a long distance from us or a much denser interstellar surrounding.
The observation of this source shows a high PF ($\sim18\%$, \citealt{kap05}) that leads to large fitting errors, and the gravitational redshift of this source is $\sim0.16$ which demonstrates $(M/M_{\odot})/(R/1\,\rm{km})\sim0.87$ indicating a stiff equation of state of NS \citep{pot16}.

\begin{figure*}
\centering
\includegraphics[width=1\textwidth]{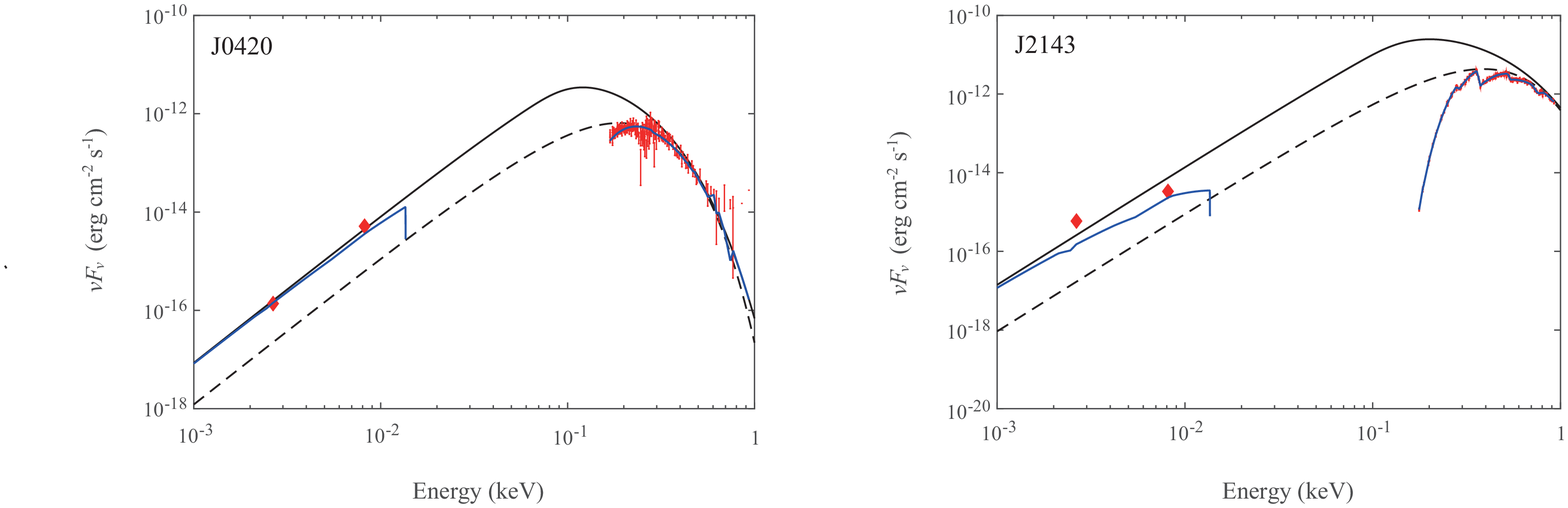}
\caption{\small{Same as Figure \ref{fig3} but for J0420(left) and J2143(right). The uniform radiative model is the black solid line.}}
\label{fig4}
\end{figure*}

The faintest X-ray source among the Seven, J0420, shows a steeper spectrum ($\beta=2.20\pm0.22$ which is greater than the case of R-J regime) at optical/UV bands that is different with other XDINSs.
This large photo index may result from the large value of PF (a factor of $\sim13\%$) which is revealed by fitting a sine wave to the X-ray pulse profile \citep{hab04}.
The data and spectral fitting by the uniform atmosphere model of J0420 and J2143 are reproduced and plotted in Figure \ref{fig4}.
In addition, J2143 shows a flatter optical spectrum ($\beta=0.53\pm0.08$).
One possible reason for the flat spectra is that a strong magnetic field ($\gtrsim10^{15}$\,G, e.g., \citealt{pon09}) blocks atmospheric matters making them diffuse hardly.
In this case, the temperature gradient should be considered (similar like a multi-color blackbody spectrum of a black hole accretion disk which presents a spectral index $\sim0.33$ at low energy bands).
Or there may be a high toroidal magnetic field that blocks transmission of heat \citep{gep06}.
Almost all of the accreted matter cannot spread out from the small pole regions that also demonstrate an extremely thin atmosphere.
We use the uniform radiative model with an absorption edge and a Gaussian absorption to fit the X-ray data of J2143.
The best fitting result shows two absorptions $E_{\rm edge}=365.6\pm2.2$\,eV with $\tau=1.07\pm0.07$ and $E_{\rm line}=759.0\pm11.1$\,eV.
These absorptions may predict a strong magnetic field.

Five of the Seven spectra could be well fitted by the radiative model, from optical/UV to X-ray bands, exhibit absorption lines (discussions of these lines can be seen in \citealt{wang17}).
In addition, we plot the absorbed bremsstrahlung and blackbody radiations to demonstrate the optical/UV excess of XDINS (see Figure \ref{fig3}).

\begin{table*}
\begin{center}
\caption{The parameters obtained from X-ray spectral fitting}
\begin{tabular}{ccccccccccc}
\hline \hline
Source RX & $kT_{\mathrm{e}}$ & $R^\infty_{\mathrm{opt}}$ & $y$ & $E_{{\rm line}}$ & $N_{\mathrm{H}}$ & $d$ & $\xi$ & $\gamma$ & $\chi^2/\mathrm{dof}$\\
 & (eV) & (km) & ($\times10^{42}\,\mathrm{cm^{-6}keV}$) & (eV) & $(\times10^{20}\,\mathrm{cm^{-2}})$ & (pc) & ($\times10^2$) & & \\
\hline
J0420.0$-$5022 & $71.6\pm2.5$ & $9.3\pm0.3$ & $1.49\pm0.25$ & 250 & $1.60\pm0.47$ & 345 & - & - & $1.15/131$\\
J0720.4$-$3125 & $179.0\pm2.2$ & $13.5\pm0.1$ & $37.3\pm2.21$ & $217.8\pm20.3$ & $1.02\pm0.33$ & 360 & 2.5 & 5 & $1.43/199$\\
J0806.4$-$4123 & $165.8\pm5.8$ & $4.5\pm0.2$ & $5.00\pm0.82$ & $445.8\pm4.3$ & $3.26\pm0.19$ & 250 & 1.5 & 3 & $1.01/423$\\
J1308.6$+$2127 & $131.4\pm2.4$ & $12.0\pm0.2$ & $10.46\pm1.13$ & $408.9\pm3.4$ & $8.77\pm0.19$ & 500 & 0 & - & $1.05/285$\\
J1605.3$+$3249 & $181.1\pm6.5$ & $13.9\pm0.5$ & $1.12\pm0.18$ & $445.2\pm4.7$ & $3.28\pm0.18$ & 390 & 5.0 & 3 & $1.12/320$\\
J1856.5$-$3754 & $96.2\pm1.1$ & $12.8\pm0.1$ & $2.11\pm0.15$ & $110.5\pm56.9$ & $0.68\pm0.26$ & 160 & 0 & - & $1.13/297$ \\
J2143.0$+$0654 & $158.0\pm3.7$ & $11.7\pm0.2$ & $10.37\pm1.29$ & $759.0\pm11.1$ & $8.36\pm2.41$ & 430 &$>1$ & $>1$ & $1.14/225$\\
\hline \hline
\end{tabular}
\end{center}
\label{tab2}
{\bf Notes.} Columns $1-8$ are source name (``Source RX''), temperature of electrons (``$kT_{\mathrm{e}}$''), radius of stars (``$R_{\rm opt}^\infty$''), parameter $y$, energy of Gaussian absorption lines, neutral hydrogen column density (``$N_{\rm H}$''), distance (``$d$''), $\xi$, $\gamma$ and $\chi^2/$degree of freedom. Errors on the spectral model parameters are derived for a $90\%$ confidence level. The distances are from \cite{kap09}. The absorption line for J0420 is fixed. J2143 also shows an edge at $E_{\rm edge}=365.6\pm2.2$\,eV with $\tau=1.07\pm0.07$.
\end{table*}

\section{DISCUSSION}
\subsection{The accretion and evolution of the XDINS}\label{4.4}
XDINSs can accrete matter from fall-back disks. The fall-back disk may be fed by ISM-matter then becomes a so-called ISM-fed debris disk accretion (IFDA). In a generalized IFDA model, NS accretes from a thin fall-back disk in the early phase through the propeller mechanism and transform to a thick-disk accretion or spherical accretion in the late phase as the disk depletes and the ISM dominate the accretion.
The radiation is supposed to come from the ISM-fed debris disk accreted matter on the stellar surface instead of the disk.
As a magnetic and accreting NS, the X-ray luminous accreted matter is a small fraction of the disk incoming matter.
Thus, the accretion rate at Alfv$\rm\acute{e}$n radius $\dot M_{\rm A}$ should be larger than the X-ray luminous accretion rate $\dot M_{\rm X}\sim10^{10}-10^{11}\,\rm{g\,s^{-1}}$ by a factor of $10-100$ \citep{tor01}.
With a weak magnetic field of $10^8\,\rm{G}\lesssim B\lesssim10^{10}\,\rm{G}$, the Alfv$\rm\acute{e}$n radius $R_{\rm A}=(\frac{R^6B^2}{\dot M_{\rm A}\sqrt{2GM}})^{2/7}\lesssim10^{10}$\,cm and the Bondi radius $R_{\rm B}\sim10^{12}$\,cm can be calculated \citep{wang17}.
For such slowly rotating NS, the light cylinder radius is calculated to be $R_{\rm LC}\sim10^{10}$\,cm. In case of $R_{\rm A}<R_{\rm LC}<R_{\rm B}$, the inflow matter would be accreted along the magnetic field lines to near polar caps and observed in simulations of disk accreted \citep{long07,rom17}.
The relationship of $\dot M_{\rm A}>\dot M_{\rm X}>\dot M_{\rm B}$ implies that the inflow matters are sufficient to maintain the radiation.
The disk mass can be estimated,
\begin{equation}\label{disk}
M_{\rm D}=\int^{R_{\rm B}}_{R_{\rm A}}2\pi \Sigma r^2 dr\sim10^{-10}M_{\bigodot},
\end{equation}
where $\Sigma$ is the standard disk density.
The disk mass is much smaller than a typical initial mass $0.1M_{\odot}$ that implies the disk has experienced a long time depleting.
In the late phase of IFDA model, the X-ray luminous accretion rate is around the Bondi accretion rate $\dot M_{\rm X}\simeq\dot M_{\rm A}\simeq\dot M_{\rm B}\simeq10^8\,\rm{g\,s^{-1}}$.
As results, isolated NSs at the late phase IFDA exhibit X-ray luminosities of $\sim10^{29}\,\rm{erg\,s^{-1}}$ that are too faint to be detected unless they are at nearby distances.
Detections of these NSs are expected from future high-resolution X-ray surveys and gravitational microlensing.

Considering gravitational capture at $R_{\rm A}$, the luminosity at $R_{\rm A}$ is $L_{\rm A}=(3/2)GM\dot M_{\rm A}/R_A=10^{27-28}\,\rm{erg\,s^{-1}}$.
The effective temperature $T_{\rm A}=[L_{\rm A}/(4\pi R_{\rm A}^2\sigma)]^{1/4}\sim0.1$\,eV, which is estimated by gravitational capture, is an upper limit.
In this case, the contribution from the disk to the optical/UV emission (X-ray as well) can be ignored.
Also, the X-ray absorption is supposed to be neglected because the density might be very low in the geometrically thick depleting disk.

With the magnetospheric radius $\sim R_{\rm A}$ is larger than the corotation radius $R_{\rm Co}=(GM/\Omega)^{1/3}\sim10^8$\,cm, the NS is in a so-called ``propeller'' stage \citep{dav73,ill75}.
Thus, one can obtain the braking torque \citep{men99,cha00}
\begin{equation}\label{1}
N=2\dot M_{\rm A}R_{\rm A}^2\Omega_{\rm K}(R_{\rm A})[1-\frac{\Omega}{\Omega_{\rm K}(R_{\rm A})}],
\end{equation}
where $\Omega$ is the angular frequency of NS and $\Omega_{\rm K}(R_{\rm A})=\sqrt{GM/R_{\rm A}^3}$ is the Kepler angular frequency at $R_{\rm A}$.
Therefore, the estimated spin-down rate $\dot P$ $\sim10^{-14}\,\rm{ss^{-1}}$ is consistent with the observation.

\subsection{Nonuniformity of the atmosphere} \label{4.1}

In Section \ref{sec2.2}, we propose that a nonuniform atmosphere would lead to the spectral index deviation.
Before that, we tried to use a distribution of cos function (i.e., Goldreich-Julian distribution) to fit the spectrum.
The reason why a cos function failed is that the radius of polar cap is very large so that there are too many higher energy photons.
If the deviation is originated from the temperature gradient (e.g., two component of blackbody), there would be a lower temperature $T\lesssim10^5$\,K component that deviates from the calculation of a low thermal conductivity on a star surface.
One possible reason is a high crustal toroidal magnetic field $B_{\rm tor}\sim10^{15}$\,G that blocks the heating transform \citep{gep06}.
In terms of the nondeviate optical spectrum (i.e., R-J spectrum), the observed optical depth $\tau_{\infty}\gtrsim1$ at $\sim10$\,eV that leads to the surface number density $n_{\rm i0}(\theta)\gtrsim10^{17}\,{\rm cm^{-3}}$.
In fact, falling atoms move along the magnetic field lines to near polar caps, and collisions between the ions and electrons may make them diffuse across the magnetic field lines.
The atmosphere would be spherically symmetrical (i.e., uniform) if these charged particles could quickly spread around the whole surface.

In case of the Seven rotate slowly, the diffusion of these charged atmospheric particles can be regarded as one-dimensional on the stellar surface.
The timescale of the collision between an ion and an electron is
\begin{equation}\label{9}
t_{\rm ie}=\frac{3(kT_{\rm e})^{1.5}m_{\rm e}^{0.5}}{4\sqrt{2\pi}e^4 n_{\rm e}\mathrm{ln}\it\bar{\Lambda}}\sim 10^7\frac{1\,\mathrm{cm^{-3}}}{n_{\rm e}}\,\mathrm{s},
\end{equation}
where $\mathrm{ln}\it\bar{\Lambda}$ is a factor of $\sim10$ for the plasma atmosphere.
Accounting the diffusion of the plasma, we have
\begin{equation}\label{10}
-\nabla p-n_{\rm e} e\bm{u}\times\bm{B}-\frac{m_{\rm e}n_{\rm e}\bm{u}}{t_{\rm ie}}=m_{\rm e}n_{\rm e}\frac{d\bm{u}}{dt},
\end{equation}
where $p$ is the pressure of the plasma, $\bm{u}$ is the velocity of electrons and $\bm{B}$ is the local magnetic field in a vertical direction.
From Equation (\ref{10}) the current density can be regarded as
\begin{equation}\label{12}
\bm{J}=n_{\rm e}\bm{u}=-\frac{D}{R}\frac{\partial n_{\rm e}}{\partial\theta}\bm{\hat{\theta}},
\end{equation}
where $D$ is defined as the diffusion coefficient and $\hat\theta$ the local orthogonal unit vectors in the directions of increasing.

Actually, there must be considered the ambipolar diffusion effect because the mass of ions is greater than that of electrons.
For the two-temperature atmosphere when $B>10^8$\,G, the ambipolar diffusion coefficient
\begin{equation}\label{13}
D_{\mathrm{A}}=(1+\frac{T_{\rm i}}{T_{\rm e}})\frac{kT_{\rm e}t_{\rm ie}}{m_{\rm e}(1+\omega_c^2t_{\rm ie}^2)}\sim10^{4}\frac{n_{\rm e}}{1\,\mathrm{cm^{-3}}}(\frac{{1\,\rm G}}{B})^2,
\end{equation}
where $\omega_{c}$ is Larmor frequency.
Diffusion is blocked by a Lorentz force with a strong magnetic field near the polar cap.
However, regarding as a dipole magnetic field, with the increase of $\theta$, collision between ions and electrons may play a leading role in blocking the diffusion against the magnetic field.
Certainly, there are still small parts of none-strange accreted matter that could permeate into the interior of the star, i.e., penetrating the strangeness barrier.
The penetration timescale of ions when there is a stable equilibrium between accretion and permeation for these falling ions is $\tau_{\mathrm{p}}\sim\Delta M/\dot{M}_{\mathrm{X}}\sim0.1-10^5$\,s \citep{wang17}.
As the electrons diffuse across magnetic field lines, spreading over almost the entire stellar surface with a low penetration, a stable diffusion equation can be described as
\begin{equation}\label{7}
\frac{\partial n_{\rm e}}{\partial t}=\frac{D_{\mathrm{A}}}{R^2}\frac{1}{\sin\theta}\frac{\partial}{\partial\theta}(\sin\theta\frac{\partial n_{\rm e}}{\partial\theta})-\frac{n_{\rm e}}{\tau_{\mathrm{p}}}.
\end{equation}
The solution of $n_{\rm e}$ can be well fitted by a single power law distribution at high latitude regions.
It is advisable that we use the toy model instead of a complete numerical solution of equation (\ref{10}).
The nonuniform toy model may predict multipole magnetic fields would be on the surface.
Then, the accreted particles moves along with this magnetic fields into high latitude regions from near polar caps.

Here, a typical diffused length of the particles which can diffuse over the whole stellar surface would be
\begin{equation}\label{15}
L=\sqrt{D_A\tau_{\mathrm{p}}}\gtrsim R.
\end{equation}
With a number density of $n_{\rm e}\sim10^{22}\,\mathrm{cm^{-3}}$ and $R\sim10$\,km, one can infer that there is a weak magnetic field ($10^{8}\mathrm{G}\lesssim B\lesssim10^{10}\mathrm{G}$) on the surface.
In fact, XDINSs are radio quiet and show purely thermal X-ray emissions that suggests their magnetospheres may be not active.
However, they are located in the upper right up of the $P-\dot{P}$ diagram ($B\sim 10^{12}-10^{14}$\,G) and beyond the death line to be ``active" NSs.
The reason why the real magnetic field is smaller than the $P-\dot{P}$-inferred magnetic field is that the propeller torque of a fallback disk may modify the period derivative \citep{liu14}.
Additionally, there may be some diffusions in the magnetosphere before accreted matter accreted into the stellar surface so that a weak magnetic field is estimated.

The X-ray optical thin atmosphere presents some faint absorption lines which may be derived from hydrocyclotron oscillation \citep{xu12} in spectra.
These absorption lines can be well fitted by single gaussian absorptions with widths $\sigma\sim0.05$\,keV, except J0720 shows a broad one ($\sigma\gtrsim1$\,keV).
We also fit the spectrum of J0720 with the nonuniform model plus a power law emission.
However, the fact of a high-energy tail is surprisingly close to a Wien spectrum \citep{vanet07} that suggests the spectrum doesn't have non-thermal (i.e., power law) components.
Therefore, this broad line may be composed of multiple absorption lines.

A normal NS atmosphere is supposed to be on top of a condensed surface \citep{ho07}.
However, the condensed surface is suggested to be maintained by a strong magnetic field ($\sim10^{13}$\,G, \citealt{lai01}).
In this case, some absorption features would be exhibited in the X-ray spectrum of J1856, but not detected with certainty.
In addition, this condensed surface described as an additional ``modified'' black body \citep{ho07}, brings more low-energy photons so that it has to introduce a large value of $N_{\mathrm{H}}$ that would consequently present a strong photoelectric absorption.
The spectral deviation, which is not an observation error, is unlikely to be originated from a uniform NS atmosphere.
And the nonuniformity of the atmosphere which is demonstrated by the X-ray pulsation, can not be understood in the frame of some creating processes of NS atmosphere (e.g., \citealt{chang03}).

\subsection{Infrared observations of XDINS} \label{4.2}
Infrared (IR) observations are very important to understand the surroundings of NS (e.g., disks around isolated NSs, \citealt{wang06,wang14}).
Accretion disks around magnetars are supported by X-ray spectral and timing observations, as well as in radio, optical, and IR observations (e.g., \citealt{tru13}).
Thus, XDINSs and magnetars which are rotating slowly as well as some high magnetic NSs are considered the most promising places to find fallback disks.
Even the luminosities of XDINSs are less than that of magnetars, XDINSs might also have a comparable near IR emission (see, Figure 2 in \citealt{mig08}).

An IR observation shows that dusty asteroid belts may surround J0806 and J2143 rather than thin dusty disks \citep{pos14}.
For the five other XDINSs, there are still not any significant evidence of warm or cold dust emission.
A possible reason for the missing disks is that they never had disks since they were born or only small dust grains could close to them just like they lost disks \citep{pos14}.
The spherical Bondi accretion can not provide a strong braking torque during the stellar evolution.
If XDINSs are during the case of a ``transitional''  disk, the IR radiation may be absorbed by the geometrically thick disk that explains the evaporated thin disk.
A dense ISM environment of $\sim10^3\,{\rm cm^{-3}}$ is ineeded.

For J2143, emission from a position of the NS is consisting of a blend of at least one southern and one brighter northern source in the red Photodetector Array Camera and Spectrometer (PACS; \citealt{pog10}) band while it is very faint emission northwest of the NS in the blue PACS band (see, Figure 5 in \citealt{pos14}).
The belt as well as these ``positional disturbances'' may result in the flat spectrum at optical bands.
\cite{lo07} shows that the IR flux upper limits derived by {\em VLT} are well above the extrapolated X-ray spectrum.
To better constrain the XDINSs' optical/NIR emission properties, there still need much deeper observations.
After around 5 years, the IR properties of XIDNS would be detected by the Large Optical Telescope (LOT)

\subsection{Geometry of XDINS} \label{4.3}

A rotating NS with nonuniform plasma atmosphere could emit pulsed emissions which would be observed.
In this case, the PF depends on the inclination angle $\alpha$ (see Figure \ref{figa1}) which tends to align in the regime of the wind braking \citep{ton17}.
It is assumed that the discovered H$\alpha$ nebula (see \citealt{van01b}) is powered by a magnetic dipole braking, a roughly NS's age of $\sim5\times10^5$ years is calculated for J1856 from its proper motion.
Thus, as shown in pulsar $P-\dot{P}$ diagram, there may be some evolution links between XDINSs and magnetars.
For instance, J1856 may be in a state of wind brake before it shows purely thermal emission \citep{ton17}.
This may indicate that XDINS may be the result of a magnetar evolution (i.e., anti-magnetar which is radio-silent emitting X-ray).

To constrain the geometry of XDINS, the two angles ($\alpha$ and $\zeta$) could be calculated through measured pulsations and polarizations.
The results of X-ray pulsation measurement are shown in Table 1 while the polarization in the soft X-rays are not feasible yet.
However, X-ray polarized properties are very significant to detect NS's magnetic field and constrain the surface (e.g., \citealt{gon16}) as well as test the model presented in this paper.
Pulse profile can also detect the particle distribution.
Moreover, both optical pulsations and polarization measurements are very difficult within reach for quite faint targets, like the Seven which have optical counterparts with magnitudes $\sim26-28$.
Only one of the seven, J1856, have been detected optical linear polarization \citep{mig17} which indicates a magnetic field $B\sim10^{13}$\,G.
While the properties of X-ray polarization may be different from that of optical polarization as a result of X-ray and optical emission come from different positions with different optical depth.
The X-ray polarised properties are expected to be tested by the Lightweight Asymmetry and Magnetism Probe (LAMP), which is supposed to work on China's space station around 2020 \citep{lamp}.

\section{SUMMARY}

In this paper, we presented a radiative model of nonuniform plasma atmosphere on an emission-negligible strangeon star's surface, and proposed that the observed emission is
the bremsstrahlung radiation from the atmosphere.
The accreted matter moves along the magnetic field lines to near polar caps, and may diffuse to other parts and makes the atmosphere nonuniform.
This allows us to understand the spectral index deviation from the expected R-J spectrum and the X-ray pulsations as well as the optical/UV excess of XDINS.
The spectra of five XDINSs (J0720, J0806, J1308, J1605 and J1856) would be well fitted in the radiative model, from X-ray to optical/UV bands, exhibiting gaussian like absorption lines.
The results of data fitting show that the electron temperatures are $\sim100-200$\,eV and that the radiation radii are $\sim5-14$\,km.

\section*{Acknowledgements}
We are grateful to Hao Tong at Guangzhou University for discussions. 
W.Y.W. and X.L.C. acknowledge the support of MoST 2016YFE0100300, Natural Science Foundation of China (NSFC) 11473044, 11633004, 11653003, CAS QYZDJ-SSW-SLH017.
Y.F. is supported by the Open Project Program of the Key Laboratory of FAST, Chinese Academy of Sciences (CAS).
X.Y.L. is supported by he West Light Foundation (XBBS-2014-23), and the National NSFC 11203018.
J.G.L is supported by the NSFC 11225314 and the Open Project Program of the Key Laboratory of Radio Astronomy, CAS.
Y.Y.L. and R.X.X. is supported by the National Key R\&D Program of China (No. 2017YFA0402602), the NSFC 11673002 and U1531243, and the Strategic Priority Research Program of CAS (No. XDB23010200).
The FAST FELLOWSHIP is supported by Special Funding for Advanced Users, budgeted and administrated by Center for Astronomical Mega-Science, Chinese Academy of Sciences (CAMS).

\bibliographystyle{mnras}
\bibliography{reference}

\section{Appendix: Flux of the bremsstrahlung from a rotating NS}
\begin{figure*}
\centering
\includegraphics[width=0.5\textwidth]{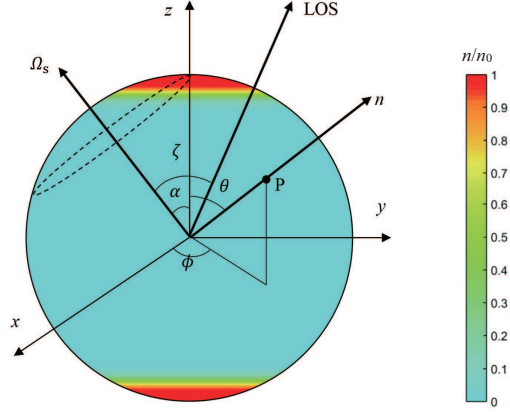}
\caption{\small{ Coordinate axes and angles used to describe the geometry and pulsed emission as well as the distribution by colour of the plasma atmosphere number density. The dashed line indicates the position of the magnetic pole $\bf{b}$ as the NS rotates about $\bf{\Omega_{\mathrm{S}}}$. The parameters of the colored electrons' number density are $\xi=150, \gamma=3$.}}
\label{figa1}
\end{figure*}

In the corotating coordinate frame ($x, y, z$), P [P=$(\theta, \phi)$, unit vector $\bf{n}$] is the emission point on the NS's surface, two angles $\zeta$ and $\alpha$ are defined: the former is the angle between the line of sight (LOS) and the spin axis $\bf{\Omega_{\mathrm{s}}}$ which is in the $\hat{x}\hat{z}$-plane, where $z$-axis parallel to the magnetic (dipole) axis, while the latter is the inclination angle that between the magnetic axis (a dipole magnetic field is assumed) and the spin axis.
Also, a fixed coordinate system is introduced, ($X, Y, Z$) with the $Z$-axis parallel to LOS and the X-axis in the LOS-spin plane.
The associated polar angles are ($\Theta_{\mathrm{S}}, \Phi_{\mathrm{S}}$) in the fixed coordinate, respectively.
The transformations linking the pairs of polar angles in the two systems are (see, \citealt{zan06})
\begin{equation}\label{a1}
\begin{cases}
\cos\theta=\bf{n}\cdot\bf{b}\\
\cos\phi=\bf{n_{\perp}}\cdot\bf{q_{\perp}},
\end{cases}
\end{equation}
where $\mathbf{n}=(\sin\Theta_{\mathrm{S}}\cos\Phi_{\mathrm{S}},\,\sin\Theta_{\mathrm{S}}\sin\Phi_{\mathrm{S}},\,\cos\Phi_{\mathrm{S}})$, $\mathbf{b}=(\sin\zeta\cos\alpha-\cos\zeta\sin\alpha\cos\omega t,\,\sin\alpha\sin\omega t,\,\cos\zeta\sin\alpha+\sin\zeta\sin\alpha\cos\omega t)$ is the magnetic axis in the fixed frame (here $\omega=2\pi/P$ is the angular velocity), an additional vector $\mathbf{q}=(-\cos\zeta\cos\omega t,\,\sin\omega t,\,\sin\zeta\cos\omega t)$, the component of $\bf{q}$ perpendicular to $\bf{b}$
\begin{equation}\label{a2}
\mathbf{q_{\perp}}=\frac{\bf{q}-(\bf{b}\cdot\bf{q})\bf{b}}{(1-\bf{b}\cdot\bf{q})^{\frac{1}{2}}},
\end{equation}
and $\mathbf{n_{\perp}}$ is defined in analogy with $\mathbf{q_{\perp}}$.

The emission from a local region of a rotating NS are shown in Figure \ref{figa1}.
Its long timescale $T$ averaged flux of the radiation could be described as
\begin{equation}\label{a3}
\bar{F}_{\nu}=\frac{1}{T}\int_{0}^{T}\mathrm{d}t\int I_{\nu}\cos\Theta_{\mathrm{S}} \mathrm{d}\Omega,
\end{equation}
where $I_{\nu}$ is the specific intensity and $\mathrm{d}\Omega$ is the solid angle in the fixed coordinate.
In the following calculations, it is assumed that the density of the strangeon star is $1.5\times2.8\times10^{14}\,{\rm cm^{-3}}$ and the atmosphere is mainly composed of hydrogens ($m_{\rm i}\sim1\,{\rm GeV}/c^2$).
The observed time averaged radiation should be gravitationally red-shifted, i.e.,
\begin{equation}\label{a4}
\bar{F}_{\nu}^{\infty}\simeq F_{\nu0}+\pi(\frac{R_{\mathrm{opt}}^\infty}{d})^2\frac{B_{\nu}}{T}\int_{0}^{T}\mathrm{d}t\int[1-\mathrm{exp}(-\tau^{\infty}_{\nu})]\mathrm{d}\Omega,
\end{equation}
where $R_{\mathrm{opt}}^{\infty}$ is the radiation radius, $d$ is the distance from the source, $B_{\nu}$ is the Planck function, $F_{\nu0}$ is the flux from a bare strangeon star which can be calculated to be
\begin{equation}\label{a5} \pi(\frac{R_{\mathrm{opt}}^\infty}{d})^2\frac{B_{\nu}}{T}\mathrm{exp}(-\tau^{\infty}_{\nu})\int_{0}^{T}\mathrm{d}t\int[1-\mathrm{exp}(-\tau^{\infty}_{\nu})]\mathrm{d}\Omega,
\end{equation}
and $\tau_{\infty}(\nu)$ is the observed optical depth at far field which can be described as
\begin{equation*}
\tau_{\infty}(\nu)=\frac{8\pi\sqrt{2}h^2n_{\mathrm{i0}}^2(\theta)e^6kT_{\mathrm{i}}}{3m_{\mathrm{e}}^{1.5}(h\nu)^{3.5}m_{\mathrm{i}}gc}[1-\mathrm{exp}(-\frac{h\nu}{kT_{\rm e}})]
\end{equation*}
\begin{equation}\label{a6}
=3.92\times10^{-45}\frac{n_{\rm i0}^2(\theta)(kT_{\rm i})_{\mathrm{keV}}}{(h\nu)^{3.5}_{\mathrm{keV}}R_{\mathrm{km}}}[1-\mathrm{exp}(-\frac{h\nu}{kT_{\rm e}})],
\end{equation}
where $h$ is the Planck constant, $m_{\rm e}$ is the mass of a electron, $T_{\rm e}$ is the temperature of electrons, $e$ is elementary charge and $c$ is the speed of light.

\end{document}